\documentclass[english,aps,prb,reprint,superscriptaddress,floatfix]{revtex4-1}
\usepackage{amsmath}
\usepackage{amssymb}
\usepackage{graphicx}
\usepackage{esint}
\usepackage{babel}
\usepackage{color}
\usepackage{bm}

\DeclareMathAlphabet{\mathcal}{OMS}{cmsy}{m}{n}

\makeatletter

\@ifundefined{textcolor}{}
{%
 \definecolor{BLACK}{gray}{0}
 \definecolor{WHITE}{gray}{1}
 \definecolor{RED}{rgb}{1,0,0}
 \definecolor{GREEN}{rgb}{0,1,0}
 \definecolor{BLUE}{rgb}{0,0,1}
 \definecolor{CYAN}{cmyk}{1,0,0,0}
 \definecolor{MAGENTA}{cmyk}{0,1,0,0}
 \definecolor{YELLOW}{cmyk}{0,0,1,0}
}

\graphicspath{{figures/}}

\begin{document}

\global\long\def\avg#1{\langle#1\rangle}
\global\long\def\p{\prime}

\global\long\def\ket#1{|#1\rangle}
\global\long\def\bra#1{\langle#1|}
\global\long\def\proj#1#2{|#1\rangle\langle#2|}
\global\long\def\inner#1#2{\langle#1|#2\rangle}
\global\long\def\tr{\mathrm{tr}}

\global\long\def\spd#1#2{\frac{\partial^{2}#1}{\partial#2^{2}}}
\global\long\def\der#1#2{\frac{d#1}{d#2}}
\global\long\def\im{\imath}

\newcommand{\<} {\left\langle}
\renewcommand{\>} {\right\rangle}
\newcommand{\dg} {\dagger}
\newcommand{\pd} {{\phantom\dagger}}

\newcommand{\ci}[1] {c_{#1}^\pd}
\newcommand{\cid}[1] {c_{#1}^\dg}

\newcommand{\ai}[1] {a_{#1}^\pd}
\newcommand{\aid}[1] {a_{#1}^\dg}
\newcommand{\jinj} {J_{\text{inj}}(\omega)}
\newcommand{\jdepl} {J_{\text{depl}}(\omega)}
\newcommand{\jinje} {J_{\text{inj}}(\epsilon_i)}
\newcommand{\jdeple} {J_{\text{depl}}(\epsilon_i)}

\newcommand{\ginj}[2] {\gamma_{#1^#2}}
\newcommand{\gk}[1] {\ginj{k}{#1}}
\newcommand{\gi}[1] {\ginj{i}{#1}}
\newcommand{\gj}[1] {\ginj{j}{#1}}
\newcommand{\hc} {\text{h.c.}}
\newcommand{\emax} {{\epsilon_{\text{max}}}}
\newcommand{\emin} {{\epsilon_{\text{min}}}}

\newcommand{\bG} {\bm{G}}
\newcommand{\bS} {\bm{\Sigma}}
\newcommand{\bGa} {\bm{\Gamma}}

\renewcommand{\Re} {\operatorname{Re}}
\renewcommand{\Im} {\operatorname{Im}}

\newcommand{\coup}{v_k}

\newcommand{\ql}{\mathcal{L}}
\newcommand{\qs}{\mathcal{S}}
\newcommand{\qr}{\mathcal{R}}
\newcommand{\qe}{\mathcal{E}}
\newcommand{\qi}{\mathcal{I}}
\newcommand{\qu}{\mathcal{U}}

% Comment this for final compile

%\newcommand{\todo}[1] {{\color{red} #1}}

% Comment/uncomment these two lines to turn section headers on/off

%\renewcommand{\section}[1] {}
%\renewcommand{\subsection}[1] {}

% These two lines show how section headers are displayed

%\renewcommand{\section}[1] {\bigskip \noindent {\large \textbf{#1}}}
%\renewcommand{\subsection}[1] {\medskip \mathversion{bold}\noindent\textbf{#1}\mathversion{normal}}

% Citation online

%\newcommand{\onlinecite}[1] {\citenum{#1}}

% Footnotes on and off

%\renewcommand{\footnote}[1] {#1}

\title{Landauer's formula with finite-time relaxation:
  Kramers' crossover in electronic transport}

\author{Daniel Gruss}

\affiliation{Center for Nanoscale Science and Technology,
             National Institute of Standards and Technology,
             Gaithersburg, MD 20899}
\affiliation{Maryland Nanocenter, University of Maryland,
             College Park, MD 20742}
\affiliation{Department of Physics, Oregon State University,
             Corvallis, OR 97331}

\author{Kirill A. Velizhanin}

\affiliation{Theoretical Division, Los Alamos National Laboratory,
             Los Alamos, NM 87545}

\author{Michael Zwolak}

\email{mpz@nist.gov}

\affiliation{Center for Nanoscale Science and Technology,
             National Institute of Standards and Technology,
             Gaithersburg, MD 20899}

\begin{abstract}
Landauer's formula is the standard theoretical tool to examine ballistic transport in nano- and meso-scale junctions, but it necessitates that any variation of the junction with time must be slow compared to characteristic times of the system, e.g., the relaxation time of local excitations. Transport through structurally dynamic junctions is, however, increasingly of interest for sensing, harnessing fluctuations, and real-time control. Here, we calculate the steady-state current when relaxation of electrons in the reservoirs is present and demonstrate that it gives rise to three regimes of behavior: weak relaxation gives a contact-limited current; strong relaxation localizes electrons, distorting their natural dynamics and reducing the current; and in an intermediate regime the Landauer view of the system only is recovered. We also demonstrate that a simple equation of motion emerges, which is suitable for efficiently simulating time-dependent transport.
\end{abstract}

%\pacs{PACS numbers: 72.10.Bg, 73.63.-b, 02.70.-c}

% General formulation of transport theory
% Electronic transport in nanoscale materials and structures
% Computational techniques; simulations

\maketitle

\section{Introduction}

The prototypical example of electron transport in nano- and meso-scale junctions is a small conducting region connected to two electron reservoirs. When the confinement in this region is strong, the rigorous treatment of quantum effects becomes crucial. The Landauer formalism \cite{landauer1957spatial, di2008electrical} is a well-known method for describing these systems, which is based on an energy-dependent transmission probability for the region of interest. This method has been successfully applied to ballistic transport, \cite{szafer1989theory} quantized conductance, \cite{van1988quantized, ohnishi1998quantized} quantum point contacts,\cite{ando1991quantum} cold-atom systems,\cite{brantut2013thermoelectric, chien2014landauer} and broadly in the area of nanoscale electronics. \cite{di2008electrical, RevModPhys.83.407, aradhya2013single} However, the viewpoint on which the Landauer formula is based neglects the explicit effect of relaxation mechanisms, the dynamics of the region of interest, and many-body interactions. Put differently, the text-book Landauer approach\cite{datta1997electronic, di2008electrical, scheer2010molecular} implicitly assumes that deviation from the equilibrium distribution in the external reservoirs is negligible. There is thus an interplay between relaxation timescales---e.g., those in the junction and at the interface with the electrodes---that cannot be captured by the Landauer formalism.

An alternative method to calculate the transport properties is to work with a closed system and explicitly solve for the  dynamics. \cite{di2008electrical, di2004transport, stefanucci2004time, bushong2005approach} This approach has been applied to study molecular conductance \cite{cheng2006simulating, evans2008spin} and induced cold atom transport, \cite{chien2012bosonic, chien2013interaction, chien2014landauer} where the latter closely approximates a closed system, giving an ideal application of this approach. The limitation of this method, however, is that the recurrence time is proportional to the total system size, meaning that a large---and computationally expensive---reservoir is needed to fully eliminate transient effects and to examine dynamical perturbations on top of an otherwise steady-state current. This is not feasible in most situations, especially if one is interested in complex, time-dependent many-body systems or the effect of relaxation (which would require the explicit incorporation of additional degrees of freedom such as phonons).

\begin{figure}
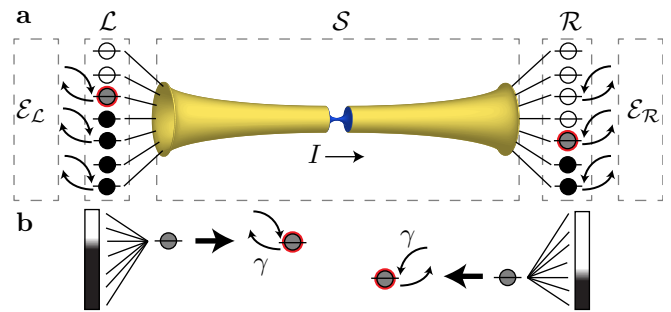

\includegraphics[width=\columnwidth]{{{fig1}}}
\caption{\textbf{Schematic representation of the model.} (\textbf{a}) System-reservoir-environment model, with yellow and blue representing the junction region (e.g., two leads connected by a junction)---the system of interest $\qs$---and $\ql$ $(\qr )$ indicating the extended reservoirs. The presence of electron sources, sinks, and interactions (electron-electron, electron-phonon, etc.), here subsumed into the environments $\qe_{\ql(\qr)}$, causes the reservoirs to relax toward their respective equilibrium distributions, which, when an external bias is applied, will be at different chemical potentials. (\textbf{b}) Each reservoir state exchanges electrons with an environment (i.e., an external reservoir at some chemical potential), which gives rise to a non-zero relaxation rate $\gamma$. The imbalance of occupied states will drive a current through $\qs$, where explicit (or implicit) relaxation mechanisms may or may not be present.} \label{fig:schematic}
\end{figure}

Transport in time-dependent structures, however, has emerged at the forefront of applications. Electronic sequencing \cite{zwolak2005electronic, lagerqvist2006fast, zwolak2008colloquium, branton2008potential, chang2010electronic, tsutsui2010identifying, huang2010identifying, ohshiro2012single} (via tunneling current through base pairs) and sensing \cite{willard2006directing, choi2012single, goldsmith2007conductance, goldsmith2008monitoring, sorgenfrei2011label, sorgenfrei2011debye, prisbrey2012electrical, sharf2012origins} (e.g., protein fluctuations on/nearby carbon nanotube and graphene devices), in particular, require a rigorous treatment of the interplay between transport and changes of the junction or molecular structure. In other words, if the local relaxation of electrons near the junction is slower or on the same timescale as the changes in the junction (typically picoseconds) the interaction between these two processes can dramatically influence transport. Practical real-time approaches to transport that are naturally suited to these systems are therefore necessary. The formalism introduced by Jauho, Meir and Wingreen \cite{meir1992landauer, jauho1994time, Haug-1996} (which has been used to describe the conduction in a variety of systems, such as quantum dots, \cite{reimann2002electronic, agrait2003quantum, baruselli2013ferromagnetic} and layered semiconductors \cite{lake1997single, platero2004photon, takei2013soft}) provides an exact formal solution to the time evolution, but involves two-time Green's functions, making its use prohibitive in many applications. In this report, we show that in the absence of time dependence, one can recover the Landauer view with reasonably sized ``extended reservoirs'' and weak relaxation. The incorporation of explicit but finite reservoirs, however, also allows for one to examine the competition between time-dependence of the junction and the relaxation rate of the reservoir region.

In transport, external sources and sinks of electrons, together with electron-electron, electron-phonon, etc., interactions, seek to sustain an equilibrium potential difference across two large regions, which we call extended reservoirs. We develop an open system approach to transport that includes a finite electron lifetime representing the presence of these relaxation mechanisms. In particular, the extended reservoirs consist of a set of states whose occupation is pushed towards equilibrium by the exchange of electrons with \emph{implicit reservoirs} (the environment $\qe$) at different chemical potentials. When a finite system is placed between them, an electric current will be driven across it. We derive a Landauer-like formula for this scenario and demonstrate that a finite relaxation time in the extended reservoirs gives rise to three distinct regimes of behavior, analogous to Kramers' turnover for chemical reactions. \cite{kramers1940brownian} A methodology similar to that below---one based on the concept of ``extended reservoirs''---was proposed and developed for classical thermal transport, \cite{velizhanin2015crossover, velizhanin2015crossover2} where a corresponding crossover effect occurs (see also Ref.~\onlinecite{biele2015timedependent}).

\section{Results}

\subsection{Model} The Hamiltonian is $H=H_{\ql}+H_{\qr}+H_{\qs}+H_{\qi}$, where the explicit degrees of freedom are divided into three parts: the left extended reservoir ($\ql$), the right  extended reservoir ($\qr$), and the system of interest ($\qs$). The extended reservoir regions have a finite electron lifetime that pushes them towards equilibrium by allowing for the exchange of electrons with the external degrees of freedom in the implicit reservoir.~\footnote{In some sense, we can say this is a grand canonical approach to transport when compared to the microcanonical approach of Ref.~\onlinecite{di2004transport}.} In other words, $\ql$ and $\qr$ are open to some larger environment $\qe$ (shown as $\qe_\ql$ and $\qe_\qr$ in Fig.~\ref{fig:schematic}), where the latter will be composed of degrees of freedom that are treated implicitly. Finally, $H_{\qi}$ describes the interaction between $\qs$ and the left ($\ql$) and right ($\qr$) extended reservoirs. Figure \ref{fig:schematic} shows a schematic of this setup.

The left and right regions each contain $N_r$ non-interacting electronic states with a Hamiltonian given by $H_{\ql} = \sum_{k\in \ql} \epsilon_{k} \cid{k} \ci{k}$ and $H_{\qr} = \sum_{k\in \qr} \epsilon_{k} \cid{k} \ci{k}$, where $k \in \ql, \qr$ indexes the single particle states and $\cid{k} (\ci{k})$ are their respective creation (annihilation) operators. The interaction Hamiltonian is described by $H_{\qi} = \sum_{k \in \ql, \qr} \sum_{i \in \qs} \hbar v_{ki} \cid{k} \ci{i} + \hc$, where $i \in \qs$ indexes the system states (with associated operators $\cid{i}, \ci{i}$). The $v_{ki}$ are the hopping rates between the reservoir and system states. The method we describe will be applicable to all dimensions, as this just changes the onsite energies in the Hamiltonian and hopping rates to the extended reservoirs. The system Hamiltonian, $H_\qs$, is arbitrary, potentially including many-body or spin-dependent interactions, vibrational degrees of freedom, etc.

% ------------------------------------------------------------------------------

In the absence of $\qs$, the extended reservoir states relax into their equilibrium occupations, i.e., their local density of electrons decays into a Fermi-Dirac distribution. The rate at which this occurs, denoted by $\gamma$, is controlled by the coupling strength between the reservoirs, $\ql$ and $\qr$, and their environment, $\qe_{\ql (\qr)}$. Generically, $\gamma$ captures the physical interaction with the environment that relax the reservoirs into equilibrium. \footnote{The relaxation rate can also be energy dependent and this will reflect both the geometry and dimensionality of the whole setup. In this work, we take $\gamma$ to be constant, independent of both $k$ and $\ql$ or $\qr$, for simplicity. This is easily relaxed, however.}  A lower bound on $\gamma^{-1}$ can be estimated by the mean scattering time in the material, which is typically on the order of $1$~fs to $10$~fs for metals. However, $\gamma$ is the  relaxation rate to reach equilibrium, which can be much weaker (especially for, e.g., local disturbances to dissipate in confined geometries, at low temperature, or in the presence of weak electron-phonon interaction). Physically, each reservoir state is exchanging electrons with a larger external reservoir ($\qe_{\ql (\qr)}$) with an applied bias of $V_{\ql(\qr)}$ and an infinite extent. The total externally applied bias is $V=V_\ql-V_\qr$, where we here take $V$ to be in units of energy. 

The general solution for the steady-states in this setup can be found by following the approach of Jauho, Meir, and Wingreen, \cite{jauho1994time} where the reservoirs are taken to be infinite with a well-defined occupation and no relaxation. Indeed, when the whole $\ql-\qs-\qr$ system is treated as some larger system $\qs^\prime$, the steady state is just the Meir-Wingreen solution, albeit with an unmanageably large number of degrees of freedom. Here, however, we are interested in calculating the transport properties of $\qs$ \emph{by itself}, i.e., to what extent can the extended reservoirs $\ql$ and $\qr$---finite in extent but with relaxation---capture the effect of infinite reservoirs in the normal approaches. As well, we want to determine what parameter ranges (e.g., realistic values of $\gamma$) are simulatable via a Markovian master equation approach. To this end, we will start with the Green's functions for the extended reservoir states uncoupled from the system, but still including a finite lifetime (note that, as we do in the Supplemental Information, one can start with all degrees of freedom treated explicitly, including $\qe_{\ql (\qr)}$, see Eq.~(A1)). For the lesser Green's function:
\begin{equation}
g^<_k(\omega) = \frac{\im \gamma f_{\ql(\qr)} (\omega)}
                     {(\omega - \omega_k)^2 + \gamma^2/4} ,
\end{equation}
with $\hbar \omega_k = \epsilon_k$ and $f_{\ql(\qr)}(\omega) = 1/(\exp[\beta(\hbar \omega -V_{\ql(\qr)})]+1)$ is the Fermi-Dirac distribution. This expression is within the wide-band approximation, see the Supplemental Information for the general case. This leads to the single particle retarded and advanced Green's functions
\begin{align}
g_k^{r(a)} (t, t') &= \mp \im \Theta (\pm t \mp t')
                              \< \{ \cid{k} (t) \ci{k} (t') \} \> \notag \\
                   &= \mp \im \Theta (\pm t \mp t')
                              e^{-\im \omega_k (t-t') -\gamma|t'-t| / 2}
\end{align}
or $g^{r(a)}_k (\omega) = (\omega - \omega_k \pm \im \gamma / 2)^{-1}$ for the Fourier transform. The $\gamma$ in both these equations reflects the finite lifetime of electrons in the extended reservoir regions. Starting with this broadened Green's function for the individual reservoir state, the steady-state current is
\begin{align} \label{eq:totalcurr}
I = \frac{e}{2\pi}\int_{-\infty}^{\infty}d\omega \;
          &\left[f_\ql(\omega)-f_\qr(\omega)\right] \notag \\
          &\times \tr \left[\bGa^{\ql}(\omega)\bG^{r}(\omega)
                            \bGa^{\qr}(\omega)\bG^{a}(\omega)\right] .
\end{align}
The quantity $\bGa_{ij}$ is the spectral density of the couplings between the system and the extended reservoirs
\begin{equation} \label{eq:spectral}
\bGa_{ij}^{\ql(\qr)}(\omega) = \sum_{k\in \ql(\qr)} v_{ik} v_{kj}
                               \frac{\gamma}{(\omega-\omega_k)^2+\gamma^2/4},
\end{equation}
with $i,j \in \qs$. $\bG^{r(a)}(\omega)$ are the exact retarded and advanced Green's functions for the system only but in the presence of the left and right extended reservoirs and including relaxation in the latter. 

Equation~\eqref{eq:totalcurr} is similar to the traditional Landauer formula. However, the density of states of a single extended reservoir state is broadened to a Lorentzian due to the inclusion of a finite relaxation time. When no interactions are present, Eq.~\eqref{eq:totalcurr} can be interpreted in terms of a relaxation-dependent transmission coefficient, $T_\gamma = \tr [ \bGa^{\ql} \bG^{r} \bGa^{\qr} \bG^{a}]$. This relaxation gives rise to different physical regimes of behavior and also an effective equation of motion that can be used to examine transport in more complex time-dependent scenarios. We first describe the different regimes of behavior for an example system.

When the reservoir states are symmetrically coupled to the system, i.e., when the distribution of energies $\epsilon_k$ and couplings $v_{ki}$ are the same for each $k\in \ql$ and its corresponding $k \in \qr$, the spectral density, Eq.~\eqref{eq:spectral}, is proportional to the imaginary part of the inverse of $\bG^{r(a)}(\omega)$. This results in a simplified expression for the current,
\begin{align} \label{eq:symcouple}
I = -\frac{e \gamma}{2 \pi}
        &\sum_{k \in \ql} \sum_{i, j \in \qs} v_{ik} v_{kj} \notag \\
        &\times \int_{-\infty}^{\infty}  d\omega
             \frac{[f_\ql(\omega) - f_\qr(\omega)]}
                  {(\omega - \omega_k)^2+\gamma^2/4}
             \Im [\bG^r_{ij}(\omega)] .
\end{align}
In the example below, we make use of this simplified expression.

% ------------------------------------------------------------------------------

\subsection{Single-Site Homogeneous System} Equation~\eqref{eq:totalcurr} is valid for any system---including those with many-body interactions---with a finite relaxation time. In what follows, however, we will focus on a homogeneous system in which the combined $\ql-\qs-\qr$ system is a 1D lattice with hopping rate $J$: $H = \sum_{n \in \ql,\qr,\qs} \hbar J (\cid{n} \ci{n+1} + \hc)$. The quantity $J$ sets the frequency scale, where the bandwidth $W = 4J$. Note that in this example, the total coupling to the system and the bandwidth are both determined by $J$. Typically, $J^{-1}$ is in the range $0.1$~fs to $1$~fs for conducting materials. We choose to work with hopping rates rather than energies as this gives more transparent expressions. 

The extended reservoir portion of $H$ can be directly diagonalized via a sine transformation. \footnote{The transformation $\qu$ is applied to the subset of states in the reservoirs, and so the couplings to $\qs$ are determined by its matrix elements.} That is, given $k \in \{1, \dots, N_r \}$, $\omega_k = -W/2 \cos [ k \pi/(N_r+1) ]$. We can express the couplings with a single index, $v_k$ for $k \in \ql, \qr$ (instead of $v_{ki}$). Using this notation, the couplings are 
\begin{equation}
v_k = J \sqrt{2 / (N_r+1)} \sin [k \pi / (N_r+1) ]. \label{eq:Coupling}
\end{equation}
Again, in this special case of a uniform 1D lattice, $J$ sets the hopping rate in both the extended reservoir region and between the system and extended reservoirs. 
Additionally, we will take the system to be a single site with no onsite energy so that $\bG^{r(a)}(\omega)=1/ \left[\omega- 2\sum_{k}v_k^2 g_k^{r(a)}(\omega) \right]$, where the sum over $k$ is in either $\ql$ or $\qr$ (the factor of 2 reflects the symmetry of the setup). 

\begin{figure}
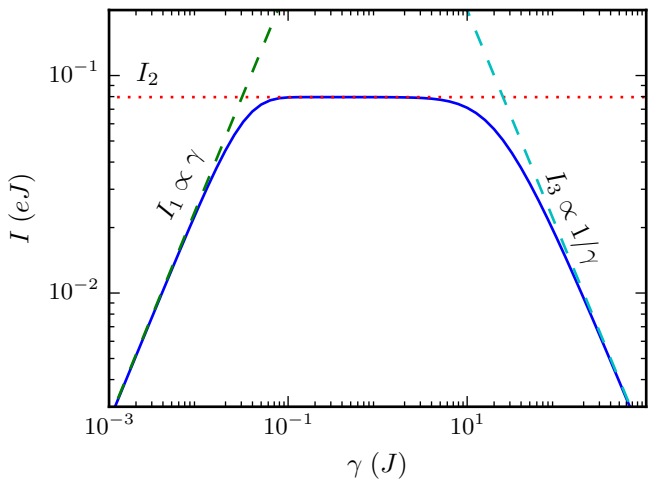

\includegraphics[width=\columnwidth]{{{fig2}}}
\caption{\textbf{Regimes of the electronic current.} The steady-state current, Eq.~\eqref{eq:symcouple} (or Eq.~\eqref{eq:totalcurr}), of the single-state system connected to two 1D extended reservoirs of size $N_r=64$. The potential difference is $V=0.5 J\hbar$ and the temperature is given by $\beta=40 (J \hbar)^{-1}$. The dashed lines show the approximations in the small and large $\gamma$ regimes, and the dotted line is the Landauer calculation of the closed system, $\qs$, with infinite $\ql$ and $\qr$ without relaxation. The small $\gamma$ regime has a current increasing linearly with $\gamma$, as it dominates the rate at which electrons flow through the whole setup. In the large $\gamma$ regime, the fast relaxation localizes electrons in the extended reservoir, causing the current to decay as $1/\gamma$. In the intermediate relaxation regime, the current matches that from a Landauer calculation.} \label{fig:linearcurrs}
\end{figure}

Figure~\ref{fig:linearcurrs} shows the calculation of the current $I$ from Eq.~\eqref{eq:symcouple} (or Eq.~\eqref{eq:totalcurr}) as a function of the relaxation rate $\gamma$ for a reservoir size $N_r=64$. There are three regimes visible: (1) a small $\gamma$ regime with current $I_1$, (2) an intermediate regime with $I_2$, and (3) a large $\gamma$ regime with $I_3$. We first discuss the intermediate regime.

\subsection{Intermediate $\gamma$} 
\begin{figure}
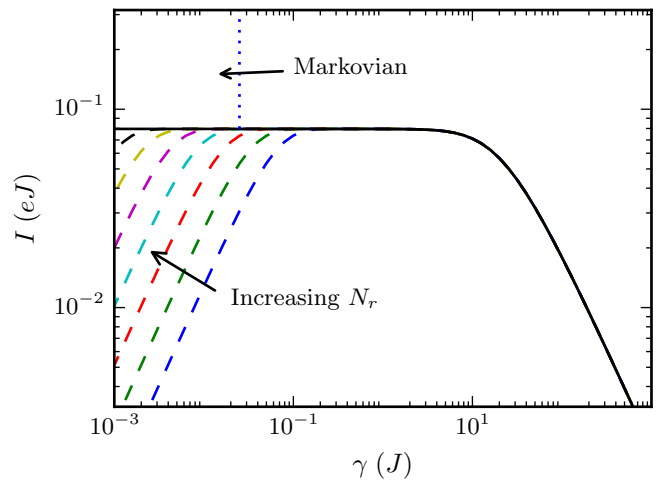

\includegraphics[width=\columnwidth]{{{fig3}}}
\caption{\textbf{Expansion of the plateau.} The steady-state current as a function of the relaxation rate $\gamma$ for the $N_r \in \{ 32, 64, 128, 256, \dots \}$ systems (dashed lines) and the $N_r \rightarrow \infty$ limit, $I_{23}$ (solid line). The parameters of the system are the same as in Fig.~\ref{fig:linearcurrs}. In the limit that $N_r \rightarrow \infty$ and then $\gamma \rightarrow 0$, we recover the standard Landauer current. The vertical dotted line demarcates the regions where the Markovian master equation is valid and not valid. Since the size of the plateau grows linearly with $N_r$, the plateau will eventually extend into the region where the Markovian equation is valid, allowing for transport in this intermediate plateau regime to be simulated with the much simpler Markovian approach.}
\label{fig:inflattice}
\end{figure}

Figure \ref{fig:linearcurrs} shows that there is an intermediate range of $\gamma$ for which the current is approximately flat. That is, in this crossover region between small and large values of $\gamma$, a plateau forms and subsequently elongates as the size of the extended reservoir increases (see Fig.~\ref{fig:inflattice}). The current in this regime is the same as that predicted by a Landauer calculation for $\qs$ alone. That calculation gives the current as $I_2=e/(2 \pi) \int_{-W/2}^{W/2} d\omega \; [f_\ql(\omega) - f_\qr(\omega)] \, T(\omega)$. In linear response, this yields
\begin{equation}
I_2 \approx eV T(\omega_F)/(2 \pi \hbar), \label{eq:LinResp}
\end{equation}
where $\hbar \omega_F$ is the Fermi level. In this example, the transmission coefficient at the Fermi level is $T(\omega_F)=1$ and the plateau comes at the quantum of conductance, $I_2(\beta \gg J \hbar) \approx eV/ (2 \pi \hbar)$ (the transmission coefficient through part of a homogeneous lattice is unity, $T(\omega)=1$, for all frequencies in the band). \footnote{At high temperature the current is $I_2(\beta \ll J \hbar) \approx e J \beta V / (2 \pi)$.} 

Perfect transmission at the Fermi level ($T(\omega_F)=1$) remains even if the hopping rate from the extended reservoir into the system is different (i.e., even if we have an inhomogeneity of the hopping rates at the interface to the system). In other systems, or in nonlinear response, though, the current---i.e., the level of the plateau---will be a complicated function of the total setup. As we discuss below, this will change when the current transitions into the other two regimes.

\subsection{Small $\gamma$} Figure \ref{fig:linearcurrs} shows that the current increases linearly with $\gamma$ when it is small. In this regime, electrons move from the left extended reservoir into the system much faster than the implicit reservoirs replenish the electrons in $\ql$ (and similarly for $\qr$). The rate of the replenishment is the relaxation rate $\gamma$, as this determines how fast the states return back to their equilibrium occupation. Thus, when $\gamma$ is small, electrons cannot be restored rapidly enough and this rate becomes the bottleneck for the current, and hence the current is essentially dependent only on $\gamma$. 

When the system is noninteracting and $\gamma$ is much less than the state spacing ($\gamma \ll W/N_r$), energy conservation guarantees that an electron coming out of state $k$ (i.e., at energy $\epsilon_k$) on the left, must exit the system at the same energy on the right. This allows the current to be broken into contributions from pairs of states, for which the pair can also be labeled by $k$ in this symmetric setup. The current flowing into the left reservoir state $k$ from the environment $\qe_\ql$ is $I_k^\ql = e \gamma (f_k^\ql - n_k^\ql)$ and the current out of that state into the system is $I^{\ql \qs}_k = e \sigma (n_k^\ql - n^\qs)$, where $f_k^{\ql(\qr)}$ is the Fermi-Dirac distribution evaluated at the reservoir state frequency $f_{\ql(\qr)}(\omega_k)$ and $\sigma$ is the particle flow rate from the reservoir state into the system. Similar rate equations hold on the right side. In the steady state ($I_k \equiv I_k^\ql = I^{\ql \qs}_k = \ldots$) and when $\gamma \ll \sigma$ (i.e., the relaxation $\gamma$ has to be weak enough that electrons are injected into an extended reservoir state much more slowly than they move into the system and are subsequently taken away from the interface between the system and extended reservoir), these equations give $I_k \approx e (\gamma /2) (f_k^\ql - f_k^\qr)$, where $\gamma/2$ is the ``reduced $\gamma$'' (i.e., it reflects that there are two interfaces, one at the left and one at the right. For different relaxation rates in $\ql$ and $\qr$, the relevant quantity would be $\gamma_\ql \gamma_\qr / (\gamma_\ql + \gamma_\qr)$). Summing over the contribution from all states $k$, the total current in this regime is 
\begin{equation} \label{eq:smallg}
I_1 \approx e \gamma / 2 \sum_k (f_k^\ql - f_k^\qr) . 
\end{equation}
The sum over $k$ is over a single set of states in the left or right, which are identical in the symmetric setup. Figure \ref{fig:linearcurrs} plots Eq.~\eqref{eq:smallg} along with the full solution, showing agreement for small $\gamma$. 

Essentially, Eq.~\eqref{eq:smallg} is just $e \gamma / 2$ times a particle bias: There are $\sum_k (f_k^\ql - f_k^\qr)$ open channels in the bias window---where an electron can move from an occupied state $k$ on the left and go to an unoccupied state $k$ on the right---and each contributes $e \gamma/2$ to the current. We note that the physics of this regime is the same as that observed in weakly coupled quantum dot systems, \cite{gurvitz1998rate} in which case $\gamma$ reflects a weak tunneling rate to the external electrodes which limits how fast the dot at the boundary can equilibrate with the electrode. 

The transition from the small to intermediate regimes occurs when the current from Eq.~\eqref{eq:smallg} intersects the plateau current, Eq.~\eqref{eq:LinResp}. We can approximate Eq.~\eqref{eq:smallg} by $e \gamma / 2 (V/\hbar) N_r /W$, where $V/\hbar$ is the bias window in terms of frequency, $W/N_r$ is the frequency spacing of the reservoir states, and, thus $(V/\hbar)/(W/N_r)$ gives the number of states in the bias window. \footnote{Note that $N_r$ should also be sufficiently large so that a significant number of states are within the bias window.} The $\gamma$ at which the transition occurs, which we will denote by $\gamma_{12}$, is 
\begin{equation}
\gamma_{12} \approx W / (\pi N_r) .
\end{equation}
This value decreases inversely with $N_r$. Indeed, as seen in Fig.~\ref{fig:inflattice}, this is responsible for the increasing size of the plateau region, as the transition to the large $\gamma$ region is independent of $N_r$ (which we will see below). 

We note that this transition $\gamma$ is equivalent to the condition necessary to be in the small gamma regime, $\gamma \ll W/N_r$. In more complex systems, or even just in nonlinear response, the transition $\gamma$ can be dependent on many other factors besides just the mode spacing, such as the hopping rate to the system, the bias, etc. In other words, the transition from small to intermediate $\gamma$ depends on the details of the setup. 

\subsection{Large $\gamma$} When $\gamma$ becomes large, Fig.~\ref{fig:linearcurrs} shows that the current ``turns over'' and starts to decay as $1/\gamma$. The strong relaxation (i.e., the fast relaxation rate) in this regime is effectively localizing electrons in the extended reservoir region. For currents to flow, electrons must remain coherent between the extended reservoir and the system. The relaxation limits this coherence to a time $\approx 1/\gamma$ and therefore the current is suppressed by this factor. 

Alternatively, this can be seen by starting with Eq.~\eqref{eq:symcouple}. There, the Lorentzian is approximately constant ($1/\gamma$) in the relevant region of integration and the Green's functions for the reservoir states, $g_k^{r(a)}$, are purely imaginary. The density of states, $\Im [ \bG^{r} ]$, is dominated by the contribution from the system in this example (see the Supplemental Information for more details). In linear response, this gives 
\begin{equation} \label{eq:largegamma}
I_3 \approx \frac{e}{2 \pi}  \left( \frac{4 \pi J^2}{\gamma} \right) ,
\end{equation}
so long as $V \neq 0$. That is, the strong relaxation renormalizes the coupling to $J^2/\gamma$ and, thus, the total electron flow through $\qs$ is limited by this factor. \footnote{This expression depends on the hopping rate to the system rather than the hopping rate in the extended reservoirs.} \footnote{As we show in the Supplemental Information, the current is related to the difference in real space occupation of the sites immediately adjacent to $\qs$ for the Markovian approach discussed below.} This also shows that the current in the large $\gamma$ regime is independent of $N_r$, with the exception of potential discretization effects (when $N_r$ is very small) that can cause mismatches in energy, and that the current is independent of the bias in this regime for the particular example we discuss.

Just as with the small $\gamma$ regime, we can find the transition into the large $\gamma$ regime. This occurs at $\gamma_{23} \approx 4 \pi J^2 \hbar / V$, where we have denoted the transition $\gamma$ as $\gamma_{23}$. Thus, while the behavior of the current in the large $\gamma$ regime is independent of bias, the transition to this regime is dependent on the bias---decreasing the bias makes this transition occur at increasingly large values of $\gamma$. 

We note that, unlike $\gamma_{12}$ and $\gamma_{23}$, how the small and large $\gamma$ behavior varies with $\gamma$ is generally independent of the form of the system and the reservoir dispersion relation, but rather only depends on characteristic quantities such as the total coupling strength (between the system and extended reservoir) and relaxation rate.

\subsection{$N_r \rightarrow \infty$ Limit} The $N_r \rightarrow \infty$ limit can be taken in Eq.~\eqref{eq:totalcurr} to regain a macroscopic electron reservoir, but with a finite relaxation time. In our example setup, the extended reservoirs become semi-infinite 1D lattices on each side. For this case, we can find the self-energy through either a recursion relation \cite{zwolak2002dna, chien2014landauer} or by integrating the states directly:
\begin{equation}
\Sigma^{r(a)}(\omega) = \sum_k \frac{v_k^2}{\omega - \omega_k \pm \im \gamma /2}
           \rightarrow
        \int \frac{  v(\omega')^2 D(\omega') d \omega'}
                  {\omega - \omega' \pm \im \gamma/2} ,
\end{equation}
with $D(\omega')= \left. dk/d \omega_k \right|_{\omega_k=\omega'}$. This expression gives a self-energy
\begin{equation} \label{eq:selfeninf}
\Sigma^{r(a)}(\omega) = \frac{8J^2}{W^2} \left( \omega \mp \im \frac{\gamma}{2} -
    \im \sqrt{\frac{W^2}{4} - \left(\omega \mp \im \frac{\gamma}{2}\right)^2 } \right) .
\end{equation}
Using this in Eq.~\eqref{eq:totalcurr} or Eq.~\eqref{eq:symcouple} (with $\bGa=2 \Im \Sigma$) provides a semi-analytic expression for the exact current through the system in the infinite $N_r$ limit, denoted by $I_{23}$. Figure~\ref{fig:inflattice} shows this quantity together with the solution for several finite $N_r$ reservoirs. \footnote{In addition, the $N_r \to \infty$ result can be expanded for small $\gamma$, yielding the lowest order contribution to the current as given by Landauer, $I_{23} \approx I_2 + O(\gamma^2)$.} These show that as $N_r$ increases the plateau will continually grow and, when $N_r \rightarrow \infty$, the small $\gamma$ regime will be eliminated entirely.

The results above are for steady-state currents, which can be calculated from exact treatment of the $\ql-\qs-\qr$ system. However, this neglects time-dependent effects present in $\qs$. As we show in the Supplemental Information, Eq.~\eqref{eq:totalcurr} also describes the steady-state solution of the Markovian master equation
\begin{align} \label{eq:master}
\dot{\rho} = -\frac{\im}{\hbar}\left[H,\rho\right]
              &+\sum_{k} \gk{+} \left( \cid{k} \rho \ci{k}
              -\frac{1}{2}\left\{ \ci{k} \cid{k}, \rho \right\} \right) \notag \\
              &+\sum_{k} \gk{-} \left( \ci{k} \rho \cid{k}
              -\frac{1}{2}\left\{ \cid{k} \ci{k}, \rho\right\} \right) 
\end{align}
in the small $\gamma$ regime and in part of the intermediate plateau region (so long as $N_r$ is sufficiently large, see Fig.~\ref{fig:inflattice}). This type of equation has been applied previously. \cite{ajisaka2012nonequlibrium, ajisaka2013nonequilibrium, zelovich2014state, ajisaka2015molecular} It is often taken as a phenomenological equation for all regimes of $\gamma$, not as a weak-coupling approximation to a memory-less reservoir. \cite{breuer2002theory} Our complete solution to both the full model (for all $\gamma$) and its Markovian counterpart enables us to put rigorous bounds on the latter's validity, which we will now discuss.

In Eq.~\eqref{eq:master}, the terms $\gk{+} = \gamma f_k^\alpha$ and $\gk{-} = \gamma (1-f_k^\alpha)$, where $\alpha=\ql(\qr)$ when $k \in \ql(\qr)$, relax the extended reservoirs into an equilibrium defined by their isolated Hamiltonian when $H_{\qi}$ is absent. That is, unlike the setup described above, this equilibrium is for the extended reservoir states at fixed energy $\hbar \omega_k$. This coincides with the concept of equilibrium above only when the broadening is sufficiently small. Larger $\gamma$, therefore, can give rise to unphysical behavior, such as residual currents at zero bias (see the Supplemental Information). In particular, the relaxation in the extended reservoirs must be smaller than the thermal relaxation, $\gamma \ll 1/\beta\hbar$ (or, in terms of timescales, $\gamma^{-1} \gg 25$~fs at room temperature), otherwise electron occupation can be smeared well above the Fermi level. As well, if one has asymmetric $\ql$ and $\qr$ extended reservoirs---with the asymmetry characterized by an energy offset $\delta$ (see the Supplemental Information)---one needs $\gamma \ll W^3 V/\delta J^2 \hbar$. Taking $\gamma$ and $N_r$ such that the current is on the plateau, $\gamma  \approx W/N_r$, gives a requirement on the extended reservoir size, $N_r \gg \delta/V$, when $\delta$ is finite. \footnote{Without an asymmetry, the anomalous current from the left reservoir to the right is canceled by the anomalous current from right to left.} This less strict condition (when compared to  $\gamma \ll 1/\beta\hbar$) guarantees that superfluous currents will be negligible compared to the actual current at finite bias. Within these regimes, the Markovian master equation allows for the calculation of the \emph{full time dynamics}. This formalism allows for the simulation of time-dependent effects or interactions and, notably, does so without the use of two-time Green's functions or the use of memory kernels, which both drastically increase the complexity of the simulations. 

\section{Discussion}

In summary, we developed the concept of extended reservoirs to examine the effect of relaxation on transport and the validity of a Markovian master equation approach. In addition to providing the full, exact solution to both the Markovian and non-Markovian cases, we showed that the current displays a crossover behavior as the relaxation rate is varied, with a weak coupling limit proportional to $\gamma$ and a strong coupling limit proportional to $1/\gamma$. These two regimes are ``relaxation'' dominated. The Landauer regime can be simulated through the use of a finite number of reservoir states and controlling the relaxation rate to be between these two regimes. The physical behavior in the presence of a finite reservoir is analogous to Kramers' problem and thermal transport.\cite{velizhanin2015crossover, velizhanin2015crossover2} 

This approach naturally leads to the Markovian master equation, Eq.~\eqref{eq:master}, for small-to-intermediate $\gamma$, which gives a suitable starting point for studying the real-time behavior of the current where the junction region is time-dependent. This formalism allows the electronic reservoirs to respond to dynamical components of the system (such as structural and energetic fluctuations) and relax back to equilibrium at a finite rate. The method, therefore, can be applied to help understand the role of fluctuations in determining transport properties, to assess the effectiveness of electronic sensing in aqueous solution, and to give a unified approach to simulating nanoscale devices out of equilibrium.

\section{Acknowledgments}

Daniel Gruss acknowledges support under the Cooperative Research Agreement between the University of Maryland and the National Institute of Standards and Technology Center for Nanoscale Science and Technology, Award 70NANB10H193, through the University of Maryland. Kirill A. Velizhanin was supported by the U.S. Department of Energy through the LANL/LDRD Program.

%\section{Contributions}
%
%M.Z. proposed the project, D.G., K.A.V., and M.Z. performed analytical calculations, and D.G. performed numerical calculations. All authors wrote the manuscript and clarified the ideas.
%
%\section{Competing Interests}
%
%The authors declare no competing financial interests.

%\bibliography{References}

%merlin.mbs apsrev4-1.bst 2010-07-25 4.21a (PWD, AO, DPC) hacked
%Control: key (0)
%Control: author (8) initials jnrlst
%Control: editor formatted (1) identically to author
%Control: production of article title (-1) disabled
%Control: page (0) single
%Control: year (1) truncated
%Control: production of eprint (0) enabled
%

\end{document}

\global\long\def\avg#1{\langle#1\rangle}
\global\long\def\p{\prime}

\global\long\def\ket#1{|#1\rangle}
\global\long\def\bra#1{\langle#1|}
\global\long\def\proj#1#2{|#1\rangle\langle#2|}
\global\long\def\inner#1#2{\langle#1|#2\rangle}
\global\long\def\tr{\mathrm{tr}}

\global\long\def\spd#1#2{\frac{\partial^{2}#1}{\partial#2^{2}}}
\global\long\def\der#1#2{\frac{d#1}{d#2}}
\global\long\def\im{\imath}

\newcommand{\<} {\left\langle}
\renewcommand{\>} {\right\rangle}
\newcommand{\dg} {\dagger}
\newcommand{\pd} {{\phantom\dagger}}

\newcommand{\ci}[1] {c_{#1}^\pd}
\newcommand{\cid}[1] {c_{#1}^\dg}

\newcommand{\ai}[1] {a_{#1}^\pd}
\newcommand{\aid}[1] {a_{#1}^\dg}
\newcommand{\jinj} {J_{\text{inj}}(\omega)}
\newcommand{\jdepl} {J_{\text{depl}}(\omega)}
\newcommand{\jinje} {J_{\text{inj}}(\epsilon_i)}
\newcommand{\jdeple} {J_{\text{depl}}(\epsilon_i)}

\newcommand{\ginj}[2] {\gamma_{#1^#2}}
\newcommand{\gk}[1] {\ginj{k}{#1}}
\newcommand{\gi}[1] {\ginj{i}{#1}}
\newcommand{\gj}[1] {\ginj{j}{#1}}
\newcommand{\hc} {\text{h.c.}}
\newcommand{\emax} {{\epsilon_{\text{max}}}}
\newcommand{\emin} {{\epsilon_{\text{min}}}}

\newcommand{\omax} {{\omega_{\text{max}}}}
\newcommand{\omin} {{\omega_{\text{min}}}}

\newcommand{\bG} {\mathbf{G}}
\newcommand{\bS} {\mathbf{\Sigma}}
\newcommand{\bGa} {\mathbf{\Gamma}}

\renewcommand{\Re} {\operatorname{Re}}
\renewcommand{\Im} {\operatorname{Im}}

\newcommand{\coup}{v_k}
\renewcommand{\sup}{L}
\newcommand{\req}{\rho_{eq}}

\newcommand{\ql}{\mathcal{L}}
\newcommand{\qs}{\mathcal{S}}
\newcommand{\qr}{\mathcal{R}}
\newcommand{\qe}{\mathcal{E}}
\newcommand{\qi}{\mathcal{I}}
\newcommand{\qu}{\mathcal{U}}

\graphicspath{{figures/}}

% Comment this for final compile

%\newcommand{\todo}[1] {{\color{red} #1}}

% Comment/uncomment these two lines to turn section headers on/off

%\renewcommand{\section}[1] {}
%\renewcommand{\subsection}[1] {}

\title{Landauer's formula with finite-time relaxation:
  Kramers' crossover in electronic transport -- Supplemental Information}

\author{Daniel Gruss}

\affiliation{Center for Nanoscale Science and Technology,
             National Institute of Standards and Technology,
             Gaithersburg, MD 20899}
\affiliation{Maryland Nanocenter, University of Maryland,
             College Park, MD 20742}
\affiliation{Department of Physics, Oregon State University,
             Corvallis, OR 97331}

\author{Kirill A. Velizhanin}

\affiliation{Theoretical Division, Los Alamos National Laboratory,
             Los Alamos, NM 87545}

\author{Michael Zwolak}

\email{mpz@nist.gov}

\affiliation{Center for Nanoscale Science and Technology,
             National Institute of Standards and Technology,
             Gaithersburg, MD 20899}

\begin{abstract}

\end{abstract}

\maketitle

\appendix

\section{Steady-State Current}

The expressions introduced in the main text are based on a derivation of the steady-state current using the equilibrium condition of a single extended reservoir site. What follows in this section is a more complete derivation of the steady-state current.

\subsection{Single Extended Reservoir State Green's Function}

Let us a consider a single electronic level of energy $\hbar\omega_k$ connected to a manifold of non-interacting states that comprise the {\em implicit} reservoir $\qe_k$. The index $k$ denotes the extended reservoir state that these states are coupled to. The $\qe_{\ql(\qr)}$ from the main text are composed of all $\qe_k$ for $k \in \ql (\qr)$. The Hamiltonian of this partial system is
\begin{equation} \label{eq:Ham_1_many}
H_k = \hbar \omega_{k} \cid{k} \ci{k}
         + \sum_{\alpha \in \qe_k} \hbar \omega_{\alpha} \cid{\alpha} \ci{\alpha}
         + \sum_{\alpha \in \qe_k} \hbar t_{\alpha}
           \left( \cid{\alpha} \ci{k} + \cid{k} \ci{\alpha} \right) .
\end{equation}
This Hamiltonian describes the systems shown in Fig. 1(b) in the main text. 
For an implicit reservoir state $\alpha$, the {\em isolated} Green's functions are
\begin{align}
g_{\alpha}^{>}(t,t')   &= -\im[1-f(\omega_{\alpha})]
                          e^{-\im\omega_{\alpha}(t-t')-\eta|t-t'|} , \\
g_{\alpha}^{<}(t,t')   &= \im f(\omega_{\alpha})
                          e^{-\im\omega_{\alpha}(t-t')-\eta|t-t'|} , \\
g_{\alpha}^{r}(t,t')   &= \theta(t-t') \left[ g^{>}(t,t')-g^{<}(t,t') \right] ,
\intertext{and}
g_{\alpha}^{a}(t,t')   &= -\theta(t'-t)\left[ g^{>}(t,t')-g^{<}(t,t') \right] ;
\intertext{or, in terms of their Fourier transforms,}
g_{\alpha}^{>}(\omega) &= -2\pi \im[1-f(\omega_{\alpha})]
                             \delta(\omega-\omega_{\alpha}) , \\
g_{\alpha}^{<}(\omega) &= 2\pi \im f(\omega_{\alpha})
                             \delta(\omega-\omega_{\alpha}) , \\
g_{\alpha}^{r}(\omega) &= 1 / (\omega-\omega_{\alpha}+\im\eta) ,
\intertext{and}
g_{\alpha}^{a}(\omega) &= 1 / (\omega-\omega_{\alpha}-\im\eta) ,
\end{align}
where $\eta$ is the infinitesimal positive number and $f(\omega)$
is the Fermi-Dirac distribution. The subscript $\alpha$ on $g_{\alpha}$ is used to distinguish it from an extended reservoir state Green's functions ($g_k$) or the full system Green's functions ($\bG_{ij}$) as used for the $\ql-\qs-\qr$ system. For the interaction of one level with all the other levels, Eq.~(\ref{eq:Ham_1_many}), we have (symbolically, on the Keldysh contour) $g_k=g_{0k}+g_{0k}\Sigma_{k}g_{k}$, with $g_{0k}$ being the single isolated {\em extended} reservoir site. Using the relation between on-contour and real-time non-equilibrium Green's functions, \cite{Haug-1996} for the retarded Green's function we have
\begin{equation}
g_{k}^{r}(\omega) = g_{0k}^{r}(\omega)
                   + g_{0k}^{r}(\omega)\Sigma_{k}^{r}(\omega)g_{k}^{r}(\omega) ,
\end{equation}
where
\begin{equation}
\Sigma_{k}^{r}(\omega)=\sum_{\alpha \in \qe_k}t_{\alpha}^{2}g_{\alpha}^{r}(\omega)
        = \frac{1}{2 \pi} \int d\omega'\,\frac{\gamma(\omega')}
                                             {\omega-\omega'+\im\eta} .
\end{equation}
Here, $\gamma(\omega)=2 \pi \sum_{\alpha}t_{\alpha}^{2}\delta(\omega-\omega_{\alpha})$. Evaluating the integral, one obtains
\begin{align}
\Sigma_{k}^{r}(\omega) = \frac{1}{2 \pi} \mathcal{P} \int d\omega'
                                 \, \frac{\gamma(\omega')}{\omega - \omega'}
                                    - \frac{\im\gamma(\omega)}{2}
                       = E_{k}(\omega) - \frac{\im\gamma(\omega)}{2} ,
\end{align}
where $E_{k}(\omega)$ is a frequency-dependent energy shift and $\gamma(\omega)$ is a frequency-dependent relaxation rate. Similarly, the lesser self-energy is evaluated as
\begin{equation}
\Sigma_{k}^{<}(\omega)=\sum_{\alpha \in \qe_k}t_{\alpha}^{2}g_{\alpha}^{<}(\omega)
                     =\im \gamma(\omega)f(\omega) .
\end{equation}
The retarded Green's function of the extended reservoir site then becomes
\begin{equation}
g_{k}^{r}(\omega) = \frac{1}{\omega-\left[\omega_{k}+E_{k}(\omega)\right]
                   + \im\gamma(\omega)/2}.
\end{equation}
As is seen, ``friction'' is in general non-Markovian, since the dephasing rate $\gamma(\omega)$ is frequency-dependent. The frequency shift, $E_{k}(\omega)$, is also non-Markovian in the same sense. Now let us find the the full lesser Green's function. To this end we will use the Keldysh equation \cite{jauho1994time,Haug-1996}
\begin{equation}
g_{k}^{<}(\omega) = g_{k}^{r}(\omega)\Sigma_{k}^{<}(\omega)g_{k}^{a}(\omega),
\end{equation}
resulting in
\begin{equation}
g_{k}^{<}(\omega) = \frac{\im\gamma(\omega)f(\omega)}
                     {(\omega-\omega_{k}-E_{k}(\omega))^{2}+\gamma^{2}(\omega)/4}.
\end{equation}
The important fact here is that the Fermi-Dirac distribution here is not evaluated at $\omega_{k}$ or at any other fixed frequency. Instead, it is evaluated at $\omega$ and, therefore, this factor is the same (at fixed $\omega$) for any Green's function of any site within the same reservoir. In particular, it guarantees that the current vanishes when the Fermi-Dirac distribution is the same for the two extended reservoirs.

The often used wide band approximation would result in $\omega$-independent $\gamma$ and vanishing $E_{k}$. In this approximation, the result for the lesser Green's function is
\begin{equation} \label{eq:gl_wide}
g_{k}^{<}(\omega)=\frac{\im\gamma f(\omega)}{\left(\omega-\omega_{k}\right)^{2}
                 + \gamma^{2}/4} .
\end{equation}
In a similar manner, the retarded Green's function is
\begin{equation} \label{eq:singlegreen}
g_{k}^{r}(\omega) = \frac{1}{\omega-\omega_{k}+\im\gamma/2} .
\end{equation}
Accordingly, one has, using the identity $g_{k}^{r}(\omega)=\left[g_{k}^{a}(\omega)\right]^{*}$,
\begin{equation} \label{eq:ident_eq}
g_{k}^{<}(\omega) = -f(\omega)\left[g_{k}^{r}(\omega)-g_{k}^{a}(\omega)\right] .
\end{equation}

\subsection{Landauer-like Formula}

We are following a notation similar to the original Meir-Wingreen paper. \cite{meir1992landauer} The Hamiltonian of the system is
\begin{equation}
H = \sum_{k\in \ql,\qr}\epsilon_{k}\cid{k}\ci{k}
    + \sum_{k\in \ql,\qr} \sum_{i \in \qs} \hbar\left(v_{ki}\cid{k}\ci{i}
                                                  + v_{ik}\cid{i}\ci{k} \right)
    + H_{\qs} ,
\end{equation}
with $\epsilon_k=\hbar\omega_k$. Unlike the Meir-Wingreen scenario where such a Hamiltonian fully describes the steady-state since the sizes of left and right reservoirs are assumed to be infinite, here we take a finite number of extended reservoir states, but each such site $k$ is connected to an implicit reservoir according to the Hamiltonian \eqref{eq:Ham_1_many}.

The time-dependent current from the left reservoir to the system can be written as \cite{meir1992landauer}
\begin{equation}
I(t) = e \sum_{k\in \ql} \sum_{j \in \qs}
          \left[v_{kj} \bG_{jk}^{<}(t,t) - v_{jk} \bG_{kj}^{<}(t,t) \right] .
\end{equation}
When the steady state is established, we can take the Fourier transform,
\begin{equation} \label{eq:Jless}
I = e \sum_{k\in \ql} \sum_{j \in \qs} \int\frac{d\omega}{2\pi} \,
       \left[v_{kj} \bG_{jk}^{<}(\omega) - v_{jk} \bG_{kj}^{<}(\omega) \right] .
\end{equation}
Since the reservoir sites are non-interacting (in a two- or more-electron sense) we have the following Dyson equation \cite{jauho1994time}
\begin{equation}
G_{kj}^{<}(\omega) = \sum_{i\in \qs} v_{ki}
     \left[g_{k}^{r}(\omega) \bG_{ij}^{<}(\omega)
           + g_{k}^{<}(\omega) \bG_{ij}^{a}(\omega) \right] ,
\end{equation}
or equivalently
\begin{equation}
G_{ik}^{<}(\omega) = \sum_{j \in qs}v_{jk}
      \left[\bG_{ij}^{r}(\omega) g_{k}^{<}(\omega)
            + \bG_{ij}^{<}(\omega) g_{k}^{a}(\omega) \right] .
\end{equation}
Using these identities, Eq.~\eqref{eq:Jless} can be rewritten as
\begin{align}
I = e \sum_{k\in \ql}\sum_{i,j \in \qs}\int\frac{d\omega}{2\pi}\, &v_{ki}v_{jk}
    \left\{ g_{k}^{<}(\omega)
           \left[\bG_{ij}^{r}(\omega)
                 - \bG_{ij}^{a}(\omega)\right] \right. \notag \\
    &\left. -\left[g_{k}^{r}(\omega) - g_{k}^{a}(\omega) \right]
           \bG_{ij}^{<}(\omega) \right\} .
\end{align}
We emphasize that even though $g_{k}(\omega)$ are single-particle non-interacting Green's functions, the corresponding quasiparticles do have finite lifetime because of their coupling to the implicit reservoirs, which is different from the original Meir-Wingreen formulation. For example, $g_{k}^{\gtrless}(\omega)$ are not delta functions with respect to $\omega$, but rather Lorentzians (when the wide band limit is taken, Eq.~\eqref{eq:gl_wide}).

To proceed further we use (i) the Keldysh equation, \cite{jauho1994time,Haug-1996} $\bG^{\gtrless}=\bG^{r}\bS^{\gtrless}\bG^{a}$, and (ii) that the self-energy due to the interaction with reservoir sites is $\bS^{r(a)}_{ij}=\sum_{k\in \ql,\qr}v_{ik}v_{kj}g^{r(a)}_{k}$, and (iii) the linear relation between real-time Green's functions $\bG^{>}-\bG^{<}=\bG^{r}-\bG^{a}$. \cite{Haug-1996} Using these identities the current can be rewritten as
\begin{widetext}
\begin{equation}
I = e \sum_{i,j \in \qs} \sum_{\alpha,\beta \in \qs}
      \sum_{k\in \ql} \sum_{l\in \ql,\qr} \int \frac{d\omega}{2\pi} \,
            v_{ki} v_{jk} v_{\alpha l} v_{l\beta}
            \bG_{i\alpha}^{r}(\omega) \bG_{\beta j}^{a}(\omega) \left\{
        g_{k}^{<}(\omega) \left[ g_{l}^{r}(\omega) - g_{l}^{a}(\omega) \right]
        - \left[g_{k}^{r}(\omega)-g_{k}^{a}(\omega)\right]g_{l}^{<}(\omega)
    \right\} .
\end{equation}
Considering the equilibrium property of the isolated state $k$, Eq.~\eqref{eq:ident_eq}, one gets
\begin{equation} \label{eq:afterequil}
I = - e \sum_{i,j \in \qs} \sum_{\alpha,\beta \in \qs}
         \sum_{k\in \ql} \sum_{l\in \qr} \int \frac{d\omega}{2\pi} \,
               v_{ki} v_{jk} v_{\alpha l} v_{l\beta}
               \bG_{i\alpha}^{r}(\omega) \bG_{\beta j}^{a}(\omega)
            \left[g_{l}^{r}(\omega) - g_{l}^{a}(\omega) \right]
            \left[g_{k}^{r}(\omega) - g_{k}^{a}(\omega) \right]
            \left\{ f_\ql(\omega) - f_\qr(\omega) \right\} .
\end{equation}
\end{widetext}

It is clearly seen that once the Fermi-Dirac distribution becomes identical on the left and on the right, the current vanishes. Actually, any partial current also vanishes, i.e., the current for a specific choice of indices $m,n,\alpha,\beta,k,l$ and frequency $\omega$, as of course is expected due to the necessary detailed balance at equilibrium.

A concise expression for the current can be written by introducing the spectral density
\begin{equation} \label{eq:specden1}
\bGa_{ji}^{\ql(\qr)}(\omega) = \im\sum_{k\in \ql(\qr)}v_{jk}v_{ki}
                       \left[g_{k}^{r}(\omega) - g_{k}^{a}(\omega) \right] ,
\end{equation}
and the expression for the current becomes
\begin{align}
I = \frac{e}{2\pi} \sum_{i,j,\alpha,\beta \in S}
    \int_{-\infty}^{\infty} d &\omega \;
    \left[f_\ql(\omega)-f_\qr(\omega)\right] \\
     &\times \bGa_{ij}^{\ql}(\omega) \bG_{j\alpha}^{r}(\omega)
             \bGa_{\alpha\beta}^{\qr}(\omega) \bG_{\beta i}^{a}(\omega) \notag .
\end{align}
Using the matrix notation,
\begin{align} \label{eq:totalcurr}
I = \frac{e}{2\pi} \int_{-\infty}^{\infty} d\omega \;
       &\left[f_\ql(\omega)-f_\qr(\omega)\right] \\
       &\times \tr \left[\bGa^{\ql}(\omega) \bG^{r}(\omega)
                         \bGa^{\qr}(\omega) \bG^{a}(\omega) \right] . \notag
\end{align}
This expression is Eq.~\eqref{main-eq:totalcurr} from the main text.

\subsection{Small $\gamma$, $N_r \rightarrow \infty$}

As an example, we examine a system that is a single noninteracting state. In this case, the continuous form for the extended reservoir self-energy from integrating the single site Green's functions, Eq.~\eqref{main-eq:selfeninf} from the main text, is
\begin{equation} \label{eq:selfeninf}
\Sigma^{r(a)}(\omega) = \left( \omega \mp \im \gamma / 2 - \im \sqrt{4J^2
                              - (\omega \mp \im \gamma /2)^2 } \right) / 2 ,
\end{equation}
when $W=4J$, and can be broken into the real and imaginary components
\begin{align}
\Re \Sigma^r(\omega) &= \frac{1}{4}\left[8 \left(\omega^2 + 4J^2\right) \gamma^2
    + 16 \left(\omega^2 - 4J^2\right)^2+\omega^4 \right]^{\frac{1}{4}} \notag \\
         &\times \sin \left[\frac{1}{2} \tan ^{-1}\left(\frac{-4 \omega \gamma}
            {-4 \omega^2 + \gamma^2 + 16J^2}\right)\right] + \frac{\omega}{2} \\
\Im \Sigma^r(\omega) &= -\frac{1}{2}
    \left[\left(\omega^2 - \frac{\gamma^2}{4} - 4J^2\right)^2
          + \omega^2 \gamma^2 \right]^{\frac{1}{4}} \notag \\
     &\times \cos \left[\frac{1}{2} \tan ^{-1}\left(\frac{-4 \omega \gamma}
             {-4 \omega^2 + \gamma^2+16J^2} \right) \right] - \frac{\gamma}{4} .
\end{align}
The term in the integral of Eq.~\eqref{eq:totalcurr} can then be written in terms of these components, and then expanded in $\gamma$ and $\omega$. Since both $\gamma$ and $\omega$ appear as the same order in the expansion and the self-energy is small outside the band edge, the error in the $\omega$ integral is small when $\gamma$ is small. In practice, the bias is typically also taken to be small. For a single central site,
\begin{align}
T_\gamma(\omega) &= \tr \left[\bGa^{\ql}(\omega) \bG^{r}(\omega)
           \bGa^{\qr}(\omega) \bG^{a}(\omega) \right] \\
&= \frac{(2\Im \Sigma^r(\omega))^2}
               {(\omega-2\Re \Sigma^r(\omega))^2 + (2 \Im \Sigma^r(\omega))^2}
        \approx 1 - \frac{\omega^2 \gamma^2}{64J^4} . \notag
\end{align}
The first term is the transmission coefficient for a 1D lattice in the Landauer formula, so the correction is of order $\gamma^2$.

\subsection{Large $\gamma$}

As $\gamma$ increases, the Green's function for a single reservoir site, Eq.~\eqref{eq:singlegreen}, approaches $g^r_k(\omega) = -2/\gamma$ in the relevant region of integration (i.e., where the difference in Fermi distributions is non-negligible). The transmission coefficient in that region, then, is 
\begin{equation}
T_\gamma(\omega) \approx - \frac{4}{\gamma} \sum_k v_k^2 \Im[\bG^r(\omega)] .
\end{equation}
Using the explicit form for the total Green's function, with $\qs$ consisting of a single non-interacting state with frequency $\omega_s$, this simplifies to
\begin{align}
T_\gamma(\omega) &\approx \frac{4}{\gamma} \sum_k v_k^2  \left( \frac{4 \sum_k v_k^2 / \gamma}{(\omega-\omega_s)^2 + (4 \sum_k v_k^2 / \gamma)^2} \right) \notag \\
&= \frac{1}{\gamma} \left( \frac{1/\gamma}{(\omega-\omega_s)^2/(4 J^2)^2 + (1/\gamma)^2} \right) \notag \\
&\approx \frac{\pi}{\gamma} \delta\left(\frac{\omega-\omega_s}{4 J^2}\right) = \frac{4 \pi J^2}{\gamma} \delta(\omega - \omega_s) ,
\end{align}
where we used that $\sum_k v_k^2 = J^2$ since the couplings $v_k$ come from a unitary transformation times the total coupling to the system. When the bias window includes this peak and at zero temperature, the current in the large $\gamma$ regime becomes
\begin{equation}
I_3 \approx \frac{2eJ^2}{\gamma} .
\end{equation}
This expression is independent of the bias, as long as it is nonzero. As we note in the main text, the transition to this value of the current, though, does depend on the bias.

\section{Markovian Master Equation}

In the main text and above, we derived a Landauer formula for the $\ql-\qs-\qr$ system, which results in a Markovian master equation for the real-time dynamics in the small-to-intermediate $\gamma$ regime. As we show below and mention in the main text, outside of this regime this formalism gives physically invalid results. In the appropriate region, though, it allows for the direct calculation of the full time dynamics and can be readily expanded to include many-body interactions or time-dependent terms. Here we will derive the full solution---in all regimes---to the Markovian master equation. Given the Hamiltonian
\begin{equation}
H = \sum_{k\in \ql,\qr} \epsilon_{k} \cid{k} \ci{k}
    + \sum_{k\in \ql,\qr} \sum_{i\in \qs} \hbar
      \left(v_{ki} \cid{k} \ci{i} + v_{ik} \cid{i} c_{k}\right) + H_{\qs} ,
\end{equation}
the starting point is the Markovian master equation,
\begin{align} \label{eq:master}
\dot{\rho} = -\frac{\im}{\hbar}\left[H,\rho\right]
             & +\sum_{k} \gk{+} \left( \cid{k} \rho \ci{k}
               -\frac{1}{2}\left\{ \ci{k} \cid{k}, \rho \right\} \right)
                                                              \nonumber \\
             & +\sum_{k} \gk{-} \left( \ci{k} \rho \cid{k}
               -\frac{1}{2}\left\{ \cid{k} \ci{k}, \rho\right\} \right) ,
\end{align}
with $\gk{+} = \gamma f_k^\alpha$ and $\gk{-} = \gamma (1-f_k^\alpha)$, where $\alpha=\ql(\qr)$ when $k \in \ql(\qr)$, and $f_k^{\ql(\qr)} = 1/(\exp[\beta(\epsilon_k-V_{\ql(\qr)})]+1)$ is the Fermi-Dirac distribution. The total applied bias is $V=V_{\ql}-V_{\qr}$. This form for $\gk{+}$ and $\gk{-}$ ensures that in the absence of the interaction $H_{\qi}$, the extended reservoirs will relax into an independent equilibrium of their own Hamiltonians, $H_\ql$ and $H_\qr$. For simplicity, we have assumed that $\gamma$ is the same in both the left and right regions, and for all $k$. 

This master equation describes the evolution of a system in the presence of explicit reservoir states, with a different mechanism than used in the main text. The following will examine the behavior of Eq.~\eqref{eq:master} over the full range of the relaxation $\gamma$.

\subsection{Single Extended Reservoir State Green's Function}

In the absence of $\qs$, the extended reservoir states decay into the equilibrium state, $\exp (-\beta H_{\ql (\qr)})/Z$, i.e., the occupations decay to a Fermi-Dirac distribution as $e^{-\gamma t /2}$. This can be shown from the exact time-dependent lesser Green's function in a reservoir site uncoupled from $\qs$, which is given by
\begin{equation}
g^<_k (0, t) = \im \< \cid{k}(t) \ci{k}(0) \>
             = \im \tr \left[ \cid{k} e^{\sup t} \ci{k} \req \right] ,
\end{equation}
where $\sup \rho = d \rho / dt$ can be found from Eq.~\eqref{eq:master} and the superoperator $\sup$ is the Lindbladian. The equilibrium state of the site has filling $f_k$, so $\tr \left[ \cid{k} \ci{k} \req \right] = f_k$. In the Fock basis of a single state, the equilibrium state is
\begin{equation}
\req = \left( \begin{matrix}
1 - f_k & 0 \\
0 & f_k
\end{matrix} \right) .
\end{equation}
A Jordan-Wigner transformation maps the electron creation and annihilation operators onto spin operators, which allows us to write $\ci{k} \req = (f_k \sigma_x + \im f_k \sigma_y) / 2$. For the equation of motion, Eq.~\eqref{eq:master}, the Lindblad operator is block diagonal in the the $\sigma_x$, $\sigma_y$ subspace and $\sigma_I$, $\sigma_z$ subspace. This means we can separately solve for the dynamics using $\sup$ for these two subspaces. If we wish to calculate the action on a generic operator $O=a_0 \sigma_I + a_x \sigma_x + a_y \sigma_y + a_z \sigma_z$, for the $\sigma_x$, $\sigma_y$ subspace:
\begin{align}
\sigma_x &\frac{d a_x}{dt} + \sigma_y \frac{d a_y}{dt}
  = -\im [\omega_k \cid{k} \ci{k}, a_x \sigma_x + a_y \sigma_y]  \\
         &+ \gk{+} \left( \cid{k} (a_x \sigma_x + a_y \sigma_y) \ci{k}
                          - \frac{1}{2} \{ \cid{k} \ci{k}, a_x \sigma_x
                          + a_y \sigma_y \} \right) \notag \\
         &+ \gk{-} \left( \ci{k} (a_x \sigma_x + a_y \sigma_y) \cid{k}
                          - \frac{1}{2} \{ \ci{k} \cid{k}, a_x \sigma_x
                          + a_y \sigma_y \} \right) . \notag
\end{align}
Acting on both sides with $ (1/2) \sigma_x \, \tr$ gives $d a_x / dt = -\gamma a_x / 2 + \omega_k a_y$ with $\gamma=\gk{+}+\gk{-}$. Similarly with $(1/2) \sigma_y \, \tr$, giving $d a_y / dt = -\gamma a_y/2 - \omega_k a_x$. Solving these equations of motion we obtain
\begin{equation}
e^{\sup t} \ci{k} \req = \sigma_x f_k / 2 e^{-t \gamma / 2 + \im \omega_k t}
     + \sigma_y f_k / 2 e^{-t \gamma / 2 - \im \omega_k t} .
\end{equation}
Then acting with $c_k$ and taking the trace gives
\begin{equation}
g^<_k (0, t) = \im e^{-t \gamma / 2 + \im \omega_k t} f_k ,
\end{equation}
for $t \geq 0$. This can be readily employed to find the retarded and advanced Green's functions for the single state
\begin{equation}
g_k^{r(a)} (t, t') = \mp \im \Theta (\pm t \mp t')
                      e^{-\im \omega_k (t-t') -\gamma|t-t'| / 2} .
\end{equation}
or its Fourier transform, $g^{r(a)}_k (\omega) = (\omega - \omega_k \pm \im \gamma / 2)^{-1}$. Physically, the reservoir sites are exchanging electrons with a larger external reservoir with an infinite number of electrons and states without memory. The lesser Green's function is also found to be
\begin{align}  \label{eq:gl_mark}
g_{k}^{<}(\omega)&=-f_{k}\left[g_{k}^{r}(\omega)-g_{k}^{a}(\omega)\right] . \notag \\
                 &=\frac{\im\gamma f(\omega_k)}{\left(\omega-\omega_{k}\right)^{2}
                 + \gamma^{2}/4} .
\end{align}
This expression recovers the original lesser Green's function, Eq.~\eqref{eq:gl_wide}, except that the distribution is evaluated at the $\omega_k$ of the state, rather than being a continuum over $\omega$. As we discuss below, this has the effect of broadening the density of states after they are occupied, rather than occupying after broadening.

\subsection{Steady-State Current}

The general solution to the steady-states of the master equation (Eq.~\eqref{eq:master}) can also be found in an analogous way to Jauho, Meir, and Wingreen and follows the same process as the derivation for Eq.~\eqref{eq:totalcurr}. In this case we use the equilibrium relation from the previous section, Eq.~\eqref{eq:gl_mark}, rather than using the one that has an $\omega$-dependent distribution $f(\omega)$. In practice, this derivation is the same but with a filling, $f_k$, dependent on the reservoir energy, $\epsilon_k$.

That is, after applying the Markovian equilibrium property, Eq.~\eqref{eq:gl_mark}, the expression for the current, Eq.~\eqref{eq:afterequil}, is instead
\begin{widetext}
\begin{equation} \label{eq:markgenform}
I = -e \sum_{i,j \in S}\sum_{\alpha,\beta \in S}\sum_{k\in L}\sum_{l\in R}
     \int \frac{d\omega}{2\pi} \, v_{ki} v_{jk} v_{\alpha l} v_{l\beta}
          \bG_{i\alpha}^{r}(\omega) \bG_{\beta j}^{a}(\omega)
          \left[g_{l}^{r}(\omega) - g_{l}^{a}(\omega)\right]
          \left[g_{k}^{r}(\omega) - g_{k}^{a}(\omega)\right]
          \left\{ f_{k}-f_{l}\right\} .
\end{equation}
\end{widetext}
Again, a concise expression for the current can be written by introducing the spectral density
\begin{equation} \label{eq:specden}
\bGa_{ji}^{\ql(\qr)}(\omega) = \im \sum_{k\in \ql(\qr)} v_{jk} v_{ki}
                    \left[g_{k}^{r}(\omega)-g_{k}^{a}(\omega)\right],
\end{equation}
and the population-weighted spectral density
\begin{equation} \label{eq:popspecden}
\tilde{\bGa}_{ji}^{\ql(\qr)}(\omega) = \im \sum_{k\in \ql(\qr)}
       f_{k} v_{jk} v_{ki} \left[g_{k}^{r}(\omega) - g_{k}^{a}(\omega) \right] ,
\end{equation}
where the Fermi-Dirac distributions are included. Then, the current can be written as
\begin{align} \label{eq:marktotalcurr}
I = \frac{e}{2\pi} \int_{-\infty}^{\infty}d\omega \; \tr
    &\left[ \tilde{\bGa}^{\ql}(\omega) \bG^{r}(\omega)
            \bGa^{\qr}(\omega) \bG^{a}(\omega) \right. \\
    &\left. \qquad - \bGa^{\ql}(\omega) \bG^{r}(\omega)
                     \tilde{\bGa}^{\qr}(\omega) \bG^{a}(\omega) \right] , \notag
\end{align}

Note that, in this case, the integrand in Eq.~\eqref{eq:marktotalcurr} does not include the Fermi-Dirac distribution as a separate prefactor as in the Meir-Wingreen / Landauer formula, but rather appears as a convolution with a Lorentzian due to the inclusion of a finite relaxation time from the Markovian master equation.

When the reservoir states are symmetrically coupled to the system, a simplified expression results:
\begin{align} \label{eq:symcouple}
I = -\frac{e}{2 \pi} \sum_{k \in \ql} &\sum_{i, j \in \qs}
      v_{ik} v_{kj} (f^\ql_k - f^\qr_k)  \\
    &\times \int_{-\infty}^{\infty} \frac{ d\omega \; \gamma}
            {(\omega - \omega_k)^2+\gamma^2/4} \Im [\bG^r_{ij}(\omega)] . \notag
\end{align}
For this symmetric case, when the two extended reservoirs have the same chemical potential, the calculated current $I$ is always equal to zero. However, in the asymmetric case, Eq.~\eqref{eq:markgenform} can yield a non-zero current even with no applied bias. We will examine this fact in Appendix~\ref{sec:bounds} in order to develop a bound for when the Markovian master equation is consistent with physical expectations.

\subsection{Small and Large $\gamma$ Limits}

In similar fashion to the main text, we can derive the limiting expressions for the steady-state current in the low and high relaxation rate regimes. The spectral function for a single reservoir site connected to an implicit bath is given by
\begin{equation}
A_{k}(\omega) = \im \left[g_{k}^{r}(\omega) - g_{k}^{a}(\omega)\right]
              = \frac{\gamma}{(\omega - \omega_{k})^{2} + \gamma^{2}/4} .
\end{equation}
If $\gamma$ is very small, $A_k(\omega)$ approaches $2\pi\delta(\omega-\omega_{k})$. Then $A_{k}(\omega)f_{k} \approx A_{k}(\omega)f_{\ql(\qr)}(\omega)$, so long as $f(\omega)$ changes little over the width of $A_{k}(\omega)$, and the expression for the current can be rewritten as
\begin{align} \label{eq:landlimit}
I_1 \approx \frac{e}{2\pi}\int_{-\infty}^{\infty}d\omega \;
        &\left[f_\ql(\omega)-f_\qr(\omega)\right] \\
        &\times \tr \left[\bGa^{\ql}(\omega) \bG^{r}(\omega)
                          \bGa^{\qr}(\omega) \bG^{a}(\omega)\right] . \notag
\end{align}
This recovers Eq.~\eqref{main-eq:totalcurr} from the main text, where the spectral density and Green's functions include the relaxation rate $\gamma$. What this means is that the master equation can be used to simulate transport, provided that $\gamma$ is in a suitable range. Even the intermediate regime can be accurately simulated, so long as $N_r$ is sufficiently large.

Similarly for large $\gamma$, the integral within Eq.~\eqref{eq:symcouple} can be calculated by transforming out from the frequency domain:
\begin{align}
\int_{-\infty}^{\infty} &\frac{ d\omega \; \gamma}
                      {(\omega - \omega_k)^2 + \gamma^2/4} \bG^r_{ij}(\omega) \\
    &= 2\pi\int_{-\infty}^{\infty} dt \, e^{-\im\omega_{k}-\gamma|t|/2} \;
            \bG_{ij}^{r}(-t)
    \approx - \im \frac{4\pi}{\gamma} \delta_{i,j} , \notag
\end{align}
which applies to both interacting and non-interacting Green's functions. In the non-interacting case, this can be thought of an integration over the density of states for each site in $\qs$. This yields an expression for the current in the large $\gamma$ regime to be
\begin{equation} \label{eq:highgamma}
I_3 \approx 2e/\gamma \sum_{k \in \ql} \sum_{i \in \qs} v^2_{ki}
        (f_k^\ql-f_k^\qr),
\end{equation}
where again the sum is over just a single set of $k$ (either in the left or right extended reservoir, which are identical).

When $\qs$ consists of a single site, then the large-$\gamma$ current is found to be
\begin{equation}
I_3 \approx 2e / \gamma \sum_{k \in \ql} v_k^2 (f_k^\ql-f_k^\qr).
\end{equation}
The sum of the $v_k^2$ terms is the transformation that diagonalizes the extended reservoir's single particle Hamiltonian. \footnote{This can also be seen working with directly with a diagonalized linear reservoir when $v_k = J \qu_{k1}$, where $\qu$ is the unitary transformation that diagonalizes the single particle Hamiltonian.} Thus,
\begin{equation}
\sum_{k \in \ql(\qr)} f_k^{\ql(\qr)} v_k^2/J^2 \rightarrow n^{\ql(\qr)},
\end{equation}
where $n^{\ql(\qr)}$ is the occupation of the extended reservoir state in real space at the site immediately adjacent to the system on the left (right). This makes a correspondence with a setup with just a single extended reservoir site on each end ($N_r=1$), $I_3 \approx 2eJ^2 (n^{\ql}-n^{\qr})/\gamma$. Essentially, the sites further away in the $N_r > 1$ case are effectively decoupled from the system as any flow of electrons away from those sites is suppressed due to the strong relaxation. Therefore, the current in the large $\gamma$ regime is also independent of $N_r$ with the exception of discretization effects.

In contrast to the Landauer-like formalism, the Markovian master equation allows for an explicit derivation showing only the occupation at the boundaries matter, i.e., all other electrons are prevented from flowing to the system by the strong relaxation. This is possible due to the well-defined occupation of each reservoir site. In the full model, the states are broadened and then occupied, so there is no equivalent transformation to a single edge site occupation like there is in the Markovian case.

We can, however, go further with the Markovian procedure. Within a small bias window, the reservoir coupling is approximately constant $v_k^2 \approx J^2/N_r$, and the sum of the occupation terms is approximately the total number of states within the bias window $\sum_k (f_k^\ql-f_k^\qr) \approx V N_r/(W \hbar)$. Substituting this in, we  find $I_3 \approx eV [2 \pi J^2 / (\gamma W)] (2 \pi \hbar)$. Comparing with Eq.~\eqref{main-eq:largegamma} in the main text, the large $\gamma$ current here is additionally inversely related to the bandwidth and grows linearly with $V$. Thus, while there is similar physical behavior to the full model, the large $\gamma$ current is quantitatively very different for the Markovian master equation, which reflects its lack of validity in this regime.

\section{Bounds on the Validity of the Markovian Master Equation}
\label{sec:bounds}

This section quantifies the regimes where the master equation is physically valid. The Markovian master equation is always mathematically valid in that it gives proper quantum evolution. However, as we will see, it does not accurately represent the equilibrium state at larger values of $\gamma$, which, e.g., leads to spurious currents and a break down of detailed balance.

\subsection{Broadening and the Fermi Level}

\begin{figure}
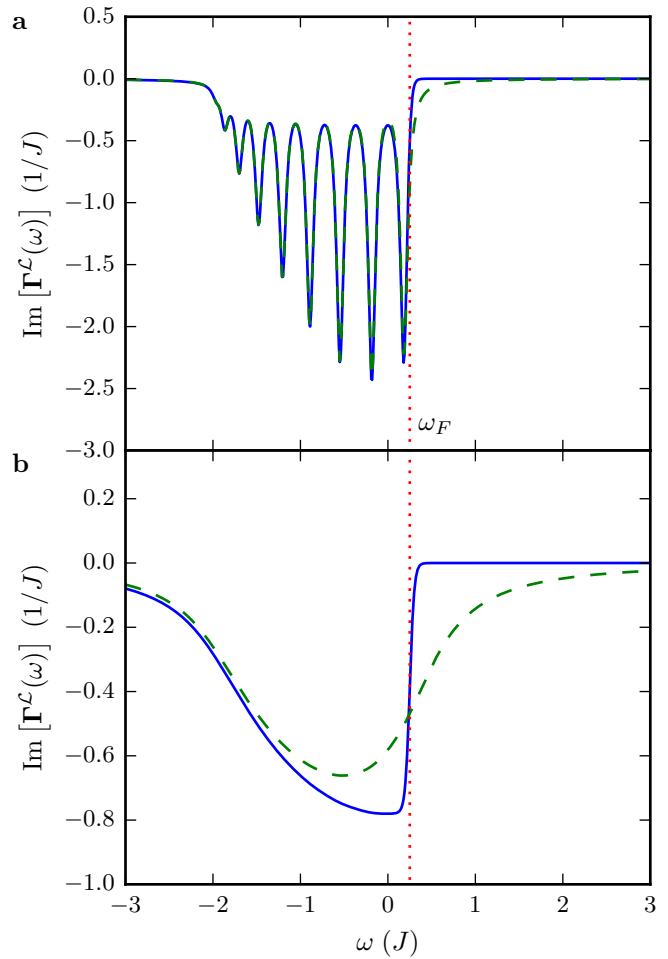

\includegraphics[width=1\columnwidth]{{{suppfig1}}}
\caption{\textbf{Spectral density and population-weighted spectral density for a 1D reservoir.} The solid line is the spectral density, Eq.~\eqref{eq:specden1} or Eq.~\eqref{eq:specden}, times the Fermi-Dirac distribution and the dashed line is the population-weighted spectral density, Eq.~\eqref{eq:popspecden}. These correspond to the Landauer-like formula and the Markovian master equation method respectively. The number of reservoir states $N_r=16$, bias and temperature are the same as the main text, $\beta=40 (J \hbar)^{-1}$ and $\omega_F = 0.25 J$, and the relaxation rate is (\textbf{a}) $\gamma=0.1 J$ and (\textbf{b}) $\gamma=J$.} \label{fig:broadfermi}
\end{figure}

The Markovian master equation broadens extended reservoir states across a wide range of $\omega$ when the relaxation rate $\gamma$ is large. That is, even with the Fermi level fixed in the isolated extended reservoir, there is excessive electron occupation beyond this level in the open system (succinctly, the Markovian equation occupies the states then broadens them, rather than broadening then occupying). As an example, Fig.~\ref{fig:broadfermi} shows the density of states times the occupation for both the full and Markovian approaches. For small $\gamma$, Fig.~\ref{fig:broadfermi}(a), the relaxation is weak enough that the states are still relatively localized in energy. However, for large $\gamma$, Fig.~\ref{fig:broadfermi}(b), the electronic occupation is smeared too much, and this allows current to flow even without a drop in chemical potential. This difference is most apparent for states at the Fermi level, $\omega_k \approx \omega_F$, as this is where the distribution is most rapidly changing. 

In the two approaches, only the lesser Green's functions are different, and these only differ by the distribution function, $f(\omega)$ compared to $f(\omega_k)$.  We quantify the error, $\Delta$, between the two by the integrated absolute difference, 
\begin{align}
\Delta &= \int_{-\infty}^{\infty} d\omega \frac{\gamma \left| f(\omega) - f(\omega_k)  \right| }{\left(\omega-\omega_{k}\right)^{2} + \gamma^{2}/4} .
\end{align}
To upper bound this error, we will use two features of $\Delta$. One is that the error is maximal when the state $\omega_k$ is at the Fermi level, $\omega_k = \omega_F$. Two is that the Fermi distribution can be replaced by the piecewise continuous function
\begin{equation}
f(\omega) \to 
\begin{cases} 
      1, & \omega \leq \omega_F - \frac{2}{\beta \hbar} \\
      \frac{1}{2} - \frac{1}{4} \beta \hbar (\omega - \omega_F), & 
           |\omega - \omega_F| < \frac{2}{\beta \hbar} \\
      0, & \omega \geq \omega_F + \frac{2}{\beta \hbar} 
   \end{cases}
\end{equation}
in order to obtain a bound of the error at the Fermi level. That is, this replacement has a greater absolute difference to $f(\omega_F)=1/2$ than the original distribution function for all $\omega$. Using these two features, the error for any $\omega_k$ is bounded by 
\begin{align}
\Delta &\le \arctan \left( \frac{-4}{\gamma \beta \hbar} \right) + \frac{\pi}{2} + \frac{\gamma \beta \hbar}{4} \ln \left( 1 + \frac{16}{\gamma \beta \hbar} \right) \\ \notag
&\lessapprox \frac{\gamma \beta \hbar}{4} \ln \left (\frac{1}{\gamma \beta \hbar} \right) .
\end{align}
The condition for this error to be small is then $\gamma \beta \hbar \ll 1$. This can be interpreted as requiring that the broadening due to the relaxation must be smaller than the broadening caused by thermal processes. As well, it has a simple, intuitive mathematical meaning: The maximal slope of the Fermi-Dirac distribution should be much smaller than $\gamma$, so that the $\gamma$-induced smearing has no significant effect on the occupation. 

Expanding on the above brief account, we can show that the error is maximal when $\omega_k$ is at the Fermi level by extremizing $\Delta$, 
\begin{equation}
\frac{d \Delta}{d \omega_k} = \frac{d}{d \omega_k} \int_{-\infty}^{\infty} d\omega \frac{\gamma \left| f(\omega) - f(\omega_k)  \right| }{\left(\omega-\omega_{k}\right)^{2} + \gamma^{2}/4} = 0 ,
\end{equation}
which can be rewritten as 
\begin{equation}
\int_{-\infty}^{\infty} d\omega \frac{\gamma \sign (\omega_k - \omega)(f'(\omega)-f'(\omega_k))}{(\omega-\omega_k)^2 + \gamma^2 / 4} = 0 .
\end{equation}
When $\omega_k$ is at the Fermi level, the term containing $f'(\omega_k)$ integrates to zero, as it is a symmetric function multiplied by an antisymmetric function around $\omega_k$. Also when $\omega_k$ is at the Fermi level, $f'(\omega)$ is symmetric around $\omega_F$ (which comes from the relationship $f(\omega_F-\omega)=1-f(\omega_F+\omega)$ when $\omega_k = \omega_F$), so $f'(\omega_F-\omega) = f'(\omega_F+\omega)$. Thus, the integrand is antisymmetric and it evaluates to zero. This therefore gives an extremum in the error. Moreover, when $\omega_k < \omega_F$, the slope is positive and when $\omega_k > \omega_F$, the slope is negative, therefore the error is maximal at the Fermi level. In the limiting cases, $\omega_k \rightarrow \pm \infty$, the total error $\Delta$ is zero, meaning that there is agreement when the state is far away from the Fermi level, as excess smearing has no consequences when the distribution is flat.

Additionally, bounds on the relaxation rate and the number of states required can be found for a given system by numerically integrating the spectral density above the Fermi level for the zero temperature case. This yields a direct measure of the improperly occupied high-energy states, even when they do not directly contribute to the electronic current.

\begin{figure}
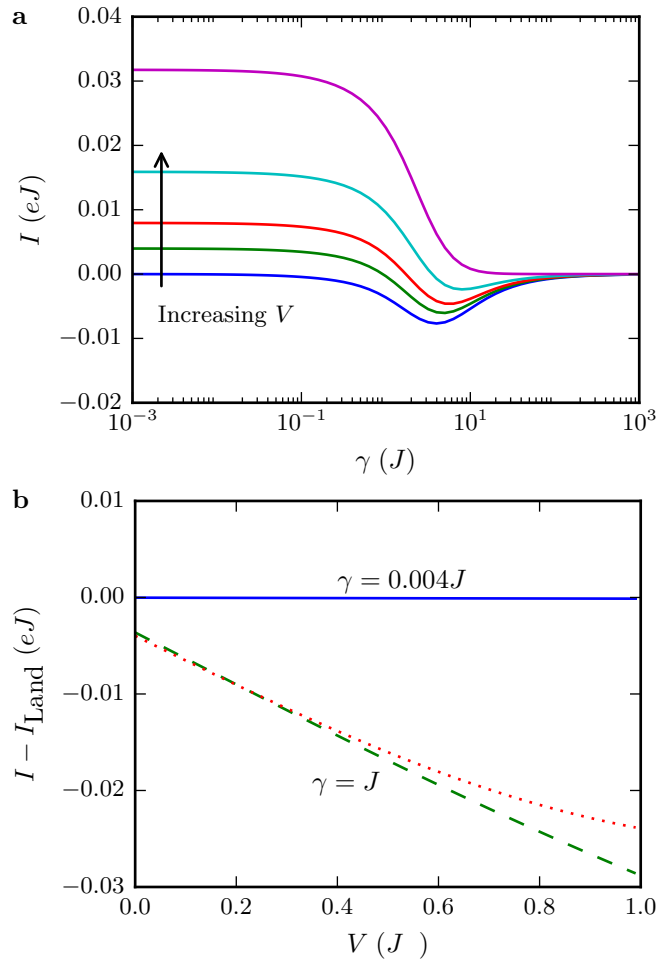

\includegraphics[width=1\columnwidth]{{{suppfig2a}}}
\includegraphics[width=1\columnwidth]{{{suppfig2b}}}
\caption{\textbf{Asymmetric system response.} (\textbf{a}) $V$ the bias dependence of current $I$ versus $\gamma$. The bias is $V \in \{ 0, 0.025, 0.05, 0.1, 0.2 \} J\hbar$ from bottom to top. The reservoirs states are equally spaced with a bandwidth of $4J$, $\beta=40(J\hbar)^{-1}$, and the left reservoir is shifted by $\delta_\ql=0.1J$ and the right is $\delta_\qr=-0.1J$, making $\delta=0.2J$. Note there is a negative current for no applied bias. (\textbf{b}) Difference between the current $I$ and the Landauer calculation $I_{\textrm{Land}}$ versus the applied bias $V$ for $N_r = \infty$. The solid line is near the peak current, $\gamma = 0.004 J$, and the dashed line is $\gamma = J$. The dotted line is the approximation from low temperature and high bandwidth Eq.~\eqref{eq:lowthighbeta} for $\gamma=J$.} \label{fig:asymcurrs}
\end{figure}

\subsection{Asymmetric Reservoirs}

Next, we will examine a system in between asymmetric reservoirs. When the asymmetry of the $\ql$ and $\qr$ extended reservoir is due to a shift in their relative energies by $\delta$, then we have $\omega_k' = \omega_k + \delta/2$ for $k \in \ql$ and $\omega_k' = \omega_k - \delta/2$ for $k \in \qr$. Solving the equation of motion, Eq.~\eqref{eq:master}, shows that there is a non-zero current for the zero applied bias ($V=0$). Note that $\delta$ is a parameter that quantifies the asymmetry of the system.

Physically, the Fermi level on both reservoirs are identical, so the steady-state current between them should be zero. When the reservoir density of states is broadened by $\gamma$, the replacement of $f(\omega)$ by $f(\omega_k)$, as the Markovian model does, results in some electronic occupation above the Fermi level. This can give rise to an electronic current. Figure~\ref{fig:asymcurrs} shows the steady-state current $I$ as a function of the relaxation $\gamma$ for increasing values of the applied bias $V$.

To simplify the analytic forms for the steady-state current, we will introduce a system where the extended reservoirs are ``Markovian'' rather than 1D. The density of states for the 1D case is approximately constant near $\omega=0$ as long as the bandwidth is large enough. For the following, the states in the reservoirs are equally spaced between frequencies $\omin$ and $\omax$ and then shifted by the asymmetry parameter $\delta$, given $k \in \{1, \dots, N_r\}$: $\omega_k = (\omin - \Delta/2) + \Delta k + \delta/2$ where the state spacing is $\Delta = (\omax - \omin) / N_r$ and the couplings are constant $v_k = \sqrt{8J^2/(2 \pi N_r)}$. Taking the limit as $N_r \rightarrow \infty$, the integral of the single particle Green's functions yields for $\ql$
\begin{equation} \label{eq:selfenbroad}
\Sigma^{r(a)}(\omega) = \frac{8J^2}{2 \pi W} \ln \left(
                \frac{\omega - \delta/2 - \omin \mp \im \gamma / 2}
                     {\omega - \delta/2 - \omax \mp \im \gamma / 2} \right) ,
\end{equation}
and similarly for $\qr$. Here the bandwidth is $W=\omax-\omin$. The calculation of the steady-state current then continues as previously.

At zero temperature ($\beta \rightarrow \infty$), expanding the current for large bandwidth (and small $V$, $\delta$), the leading terms are
\begin{align} \label{eq:lowthighbeta}
I \approx \frac{eV (8J^2/W)}{\hbar \pi [\gamma + 2 (8J^2/W)]}
          - \frac{8e \delta \gamma J^2}{\pi W^3} .
\end{align}
The first term gives the linear response current. For small $\gamma$, it is $eV/2\pi\hbar$. The second term gives a residual, unphysical current. So long as the asymmetry is small compared to the bandwidth or $\gamma$ is sufficiently small, this term will be negligible compared to the linear response contribution. Moreover, to accurately calculate results in the Landauer regime for steady states, $\gamma$ needs to be of the order of $W/N_r$ (this condition depends on the details of the setup, as discussed in the main text regarding the small $\gamma$ regime). Thus, so long as 
\begin{equation}
\gamma \approx W/N_r \ll W^3 V/\delta J^2 \hbar
\end{equation}
then the simulation will accurately predict steady state behavior. In other words, one needs $N_r \gg J^2 \delta \hbar/W^2 V$ (or, for our particular example in the main text,  $N_r \gg \delta / V$), which is a condition that is basically always fulfilled, to ensure the simulation gives accurate results in the Landauer regime. This covers both small and intermediate regimes discussed throughout the text.

\section{Linear Reservoir with Equally Spaced States}

The process of discretization of the states can have a large effect on the calculated current when using the master equation, particularly when the bias window lies completely in a gap between states. As an example, in the 1D lattice case, the extended reservoir portion of $H$ can be directly diagonalized via a sine transformation. Thus, if one wished to study a large---but still finite---extended reservoir driving a current through a time-dependent junction, one could increase $N_r$ by increasing the size of the reservoir in real space, a fact that also applies in higher dimensions. However, it can be more useful to instead take $N_r$ states evenly spaced in energy and increase $N_r$ by decreasing the spacing. That is, given $k \in \{1, \dots, N_r \}$, $\omega_k = -(W/2 + \Delta/2) + \Delta k$, where the state spacing is $\Delta = W / N_r$, places the states evenly within the energy band (note that $W=4J$ for this model).  In order for this to represent the 1D extended reservoir, the coupling constants to the system need to incorporate the local density of states for the real-space lattice at the boundary with the system. We can write the couplings with a single index, $v_k$ for $k \in \ql, \qr$ (instead of $v_{ki}$), as the coupling to the system's boundaries only depend on $k$. Using this notation, the couplings
\begin{equation}
v_k = 2 (4J^2 - \omega_k^2)^{1/4} \sqrt{J / 2 \pi N_r},
\label{eq:vkwDOS}
\end{equation}
have a factor of $\sqrt{4J^2 - \omega_k^2}$, which is the local density of states for the 1D lattice. In the $N_r \rightarrow \infty$ limit, this choice of state discretization recovers the semi-infinite, real-space lattice.

\section{Single-Site Reservoir Rate Equation}

When the broadening $\gamma$ is much smaller than the state spacing $\approx W/N_r$ and for a non-interacting system, conservation of energy requires that each electron entering the system from a site with a given energy $\epsilon_k$ also leave the system from the a site with the same energy. This allows the current to be broken into contributions from pairs of extended reservoir states, which can be calculated from a system of rate equations. We shall examine a three site system which consists of a single site from each of $\ql$ and $\qr$ (indexed by $k$) and a single site in $\qs$. Further, we will assume different relaxation rates on the left and right, $\gamma_\ql$ and $\gamma_\qr$ respectively to make the appearance of a ``reduced $\gamma$'' clear.

When $\gamma$ is small, the effect of the environment $\qe_{\ql(\qr)}$ is to relax each extended state to a target filling, denoted by $f^{\ql(\qr)}$, at a rate $\gamma_{\ql(\qr)}$. This current due to relaxation is proportional to the difference between the onsite occupation of the reservoir state, $n^{\ql(\qr)}$, and $f^{\ql(\qr)}$. In linear response, the current from $\qe_\ql$ and the current into $\qe_\qr$ are
\begin{equation}
I^\ql_k \approx e\gamma_\ql(f^\ql_k - n^\ql_k), \qquad
I^\qr_k \approx e\gamma_\qr(n^\qr_k - f^\qr_k) .
\end{equation}
The current between $\qs$ and the reservoir states is proportional to a rate parameter $\sigma$ times the difference in occupations:
\begin{equation}
I^{\ql\qs}_{k} \approx e\sigma (n^\ql_k - n^\qs) , \qquad
I^{\qs\qr}_{k} \approx e\sigma (n^\qs - n^\qr_k) .
\end{equation}
The value of $\sigma$ is related to the coupling between the reservoir and system and describes the particle flow rate. In the general case, this would be a function of the system-reservoir coupling and the total Green's function. For the example system, however, $\sigma$ scales as $1/N_r$. In linear response, the conductance is independent of $J$ and $W$, so the sum of the total current from all $N_r$ states must be a constant.

In the steady state, all four of these currents---$I^\ql_k$, $I^\qr_k$, $I^{\ql\qs}_{k}$, $I^{\qs\qr}_{k}$---must be equal, giving the solution
\begin{equation} \label{eq:ratecurr}
I_k = \frac{e \sigma}{\left(2 + \sigma \frac{\gamma_\ql + \gamma_\qr}{\gamma_\ql \gamma_\qr}\right)} (f^\ql_k - f^\qr_k) .
\end{equation}
The quantity $\gamma_\ql \gamma_\qr / (\gamma_\ql + \gamma_\qr)$ is the ``reduced $\gamma$'' between the two reservoirs and is simply $\gamma/2$ when the relaxation rates are equal. The small $\gamma$ ($\gamma \ll \sigma$) approximation for Eq.~\eqref{eq:ratecurr} with equal relaxation rates is $I_k \approx e (\gamma/2) (f_k^\ql - f_k^\qr)$, as used in the main text. 

This rate argument can be extended to the full $N_r \neq 1$ system by solving a system of $4 N_r$ equations for each of the incoming and outgoing currents and then equating the total current into $n_\qs$, $\sum_k I^{\ql\qs}_k$, with the total current leaving $\sum_k I^{\qs\qr}_k$.  When $\gamma_\ql = \gamma_\qr = \gamma$, this yields the current through the system as
\begin{equation} \label{eq:ratetotcurr}
I \approx e \gamma / 2 \sum_k (f_k^\ql - f_k^\qr) \frac{\sigma}{\sigma+\gamma} .
\end{equation}
For small $\gamma$, this recovers the expression in the main text. Additionally, the occupation of the central site, $n^\qs$, is found to be equal to the mean target filling, $\sum_k (f^\ql_k+f^\qr_k)/(2N_r)$. With a symmetrical distribution of reservoir state energies and bias window (as in the case of the main text), then the central site onsite density is $1/2$.

Lastly, a full solution involving a different $\sigma_k$ for each reservoir mode is obtainable. Assuming that the $\sigma_k$ are symmetric between the left and right sides, the total current flow recovers Eq.~\eqref{eq:ratetotcurr} to the lowest order of $\gamma$. The relaxation strength limits the total current through the system and, so long as $\gamma$ is sufficiently small, then the current is independent of the system-reservoir coupling.

%\bibliography{References}

%merlin.mbs apsrev4-1.bst 2010-07-25 4.21a (PWD, AO, DPC) hacked
%Control: key (0)
%Control: author (8) initials jnrlst
%Control: editor formatted (1) identically to author
%Control: production of article title (-1) disabled
%Control: page (0) single
%Control: year (1) truncated
%Control: production of eprint (0) enabled
%